\documentclass{iopconfser}

\usepackage{graphicx}
\usepackage{amsmath}
\usepackage[T1]{fontenc}
\usepackage{lmodern}
\usepackage{comment}
\usepackage[colorlinks=true, linkcolor=blue, citecolor=blue, urlcolor=blue]{hyperref}

\begin{document}

\title{Theory-agnostic searches of non-gravitational modes in the ringdown}

\author{Francesco Crescimbeni$^{1,2}$}

\affil{$^1$Dipartimento di Fisica, Sapienza Università di Roma, Piazzale Aldo Moro 5, 00185, Roma, Italy}
\affil{$^2$INFN, Sezione di Roma, Piazzale Aldo Moro 2, 00185, Roma, Italy}

\email{francesco.crescimbeni@uniroma1.it}

\begin{abstract}
We present a theory-agnostic framework to search for extra-fundamental modes in the ringdown phase of black hole mergers. These modes, distinct from standard General Relativity (GR) deviations, originate from modifications of the theory of gravity. Applying our method to the events GW150914, GW190521, and GW200129, we find no significant evidence for extra modes beyond those expected in GR.
\end{abstract}

\section{Introduction}

Black hole~(BH) spectroscopy~\cite{Berti:2025hly} is a prime tool for strong-field tests of GR~\cite{LIGOScientific:2021sio}, as it extracts the remnant quasinormal modes~(QNMs) \cite{Vishveshwara:1970zz,Berti:2009kk} from the ringdown, uniquely determined by the final mass and spin. This enables both null-hypothesis tests \cite{Isi:2019aib,Franchini:2023eda} and remnant characterization~\cite{Maggio:2020jml}. Extensions of GR generically introduce extra degrees of freedom~\cite{Lovelock:1971yv,Sotiriou:2014yhm}, such as scalars in scalar-tensor or Horndeski gravity~\cite{Horndeski:1974wa}, high-curvature corrections~\cite{Alexander:2009tp}, Einstein-Aether~\cite{Jacobson:2000xp}, Hořava–Lifshitz~\cite{Horava:2009uw}, or massive gravity~\cite{deRham:2010kj}. These can deform Kerr QNMs and, crucially, excite \textit{extra modes} coupled to gravity~\cite{Molina:2010fb,Pani:2013pma,Cardoso:2020nst,DAddario:2023erc, Lestingi:2025jyb}, which vanish in the GR limit, but provide clean beyond-GR signatures. 

\section{A theory agnostic test for non-gravitational modes in the ringdown}
Beyond-GR theories mentioned above motivate a new test based on a GR ringdown augmented by additional modes, as follows \cite{Crescimbeni:2024sam}:
\begin{align}\label{eq:rdmodel2}
    h(t)=&\sum_{i} A_{i}\cos\left(2\pi f_i^{\rm Kerr}(1+\delta f_i)t+\phi_{i}\right)
    e^{-\frac{t}{\tau_i^{\rm Kerr}(1+\delta \tau_i)}}+\sum_{i} \hat A_{i}\cos\left(2\pi \hat f_{i}t+\hat \phi_{i}\right)e^{-{t}/{\hat\tau_{i}}}\,.
\end{align}
Here $i\equiv(l,m,n)$ labels the multipolar, azimuthal and overtone indices; $A_i$ and $\phi_i$ are the amplitude and phase of the standard Kerr gravitational QNMs, with frequencies $f_i^{\rm Kerr}$ and damping times $\tau_i^{\rm Kerr}$, and $\delta f_i, \delta \tau_i$ the GR deviations. The hatted quantities $(\hat A_i,\hat\phi_i,\hat f_i,\hat\tau_i)$ denote the corresponding parameters of the extra modes. Standard tests of gravity based on the ringdown rely on the first term in Eq.~\eqref{eq:rdmodel2}: their primary goal is to measure $\delta f_i$ and $\delta \tau_i$ and to assess whether these quantities 
are consistent with the null hypothesis~\cite{LIGOScientific:2021sio}. In what follows, we neglect $\delta f_i$ and $\delta \tau_i$, since these parameters have already been extensively investigated in standard ringdown tests and are, in general, theory-dependent and 
challenging to compute.
 On the other hand, the frequencies and damping times of the extra modes can be expanded as~\cite{Cardoso:2009pk,Molina:2010fb,Blazquez-Salcedo:2016enn,Cardoso:2020nst}
\begin{align}
    \hat f_i=f_i^{{\rm Kerr},\,s}(1+\delta \hat f_i)\,,\qquad \hat \tau_i=\tau_i^{{\rm Kerr},\,s}(1+\delta \hat \tau_i)\,, \label{eq:QNMmod2}
\end{align}
where $f_i^{{\rm Kerr},\,s}$ and $\tau_i^{{\rm Kerr},\,s}$ are the QNMs of a test field in the Kerr metric, and also in this case $\delta \hat f_i$ and $\delta \hat \tau_i$ are complicated, theory-dependent, functions of the mass, spin, and coupling constants, while $s=0,1$ are respectively associated to scalar and vector fields. Crucially, in this case the amplitudes $\hat A_i$ of these modes are \emph{proportional} to (powers of) the coupling constants~\cite{Cardoso:2009pk,Molina:2010fb,Blazquez-Salcedo:2016enn,Cardoso:2020nst,DAddario:2023erc}. Therefore, to leading order in the corrections, we can neglect $\delta \hat f_i$ and $\delta \hat \tau_i$, so that the GR deviations are generically parametrized only by the amplitude of the \emph{test-field} modes on a GR BH background. This approach is complementary to standard spectroscopy, theory-agnostic, and akin to searches for extra GW polarizations~\cite{Isi:2017fbj,LIGOScientific:2021sio}, offering a new route to probe gravity beyond GR~\cite{Brito:2018rfr,LIGOScientific:2021sio}. 

\section{Searching for extra ringdown modes in O3 data}
\begin{figure}[t]
\centering
\includegraphics[width=0.52\textwidth]{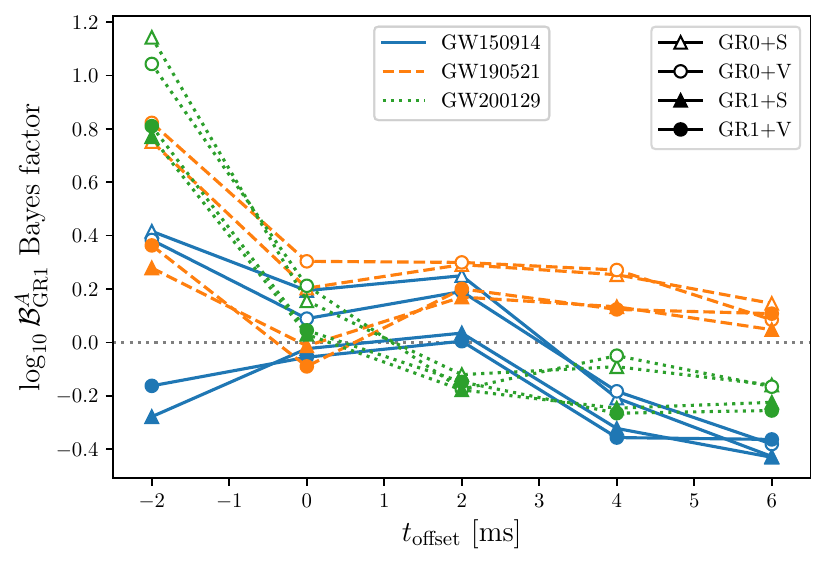}
\caption{$\log_{10}$ Bayes factors for various ringdown models with extra scalar or vector modes (labelled with 'A') relative to the \texttt{GR1} model, as a function of the offset time $t_{\rm offset}$. Different colors and line styles denote different events.}
\label{fig:BF_scalar}
\end{figure}

To compare with observational data the framework of Eq.~\eqref{eq:rdmodel2} with the constraints imposed in the previous section, we construct a set of minimal models that progressively include extra modes: 1) \texttt{GR1}, the standard GR spectrum with the fundamental $(2,2,0)$ mode and the first overtone $(2,2,1)$; 2) \texttt{GR0+S} (\texttt{GR0+V}), the fundamental GR mode $(2,2,0)$ plus an additional scalar (vector) extra mode $(2,2,0)$; 3) \texttt{GR1+S} ({GR1+V}), the GR modes $(2,2,0),(2,2,1)$ supplemented by a scalar (vector) extra field mode $(2,2,0)$. We applied this test to three GW events with significant ringdown content: GW150914~\cite{LIGOScientific:2016aoc}, GW190521~\cite{LIGOScientific:2020iuh}, and GW200129~\cite{LIGOScientific:2021sio}. We implement a Bayesian analysis for all those models by using \texttt{pycbc inference} \cite{Biwer:2018osg}. The main results are shown in Fig.~\ref{fig:BF_scalar}, which reports the Bayes factors of different models relative to \texttt{GR1}.
For all three events, $\log_{10}{\cal B}{\texttt{GR1}}^{\rm A}$ remains close to zero at all $t_{\rm offset}$, which means that models with extra scalar or vector modes are statistically indistinguishable from \texttt{GR1}. This suggests that current ringdown SNRs are not high enough to detect such modes, and could be confused with a putative detection of a gravitational overtone. The modest rise of Bayes factors at negative $t_{\rm offset}$ reflects the model beginning to absorb pre-merger and overtone contributions, rather than genuine evidence for extra modes.

\section{Conclusions}
We have introduced a theory-agnostic and practical test to search for extra ringdown modes, complementary to ordinary BH spectroscopy, that focuses on QNM shifts. Although present events lack the necessary SNR to reveal such signatures, forecasts with third-generation detectors indicate that future detectors will be sensitive to extra-mode amplitudes of $\mathcal{O}(10^{-2})$ with respect to the fundamental mode, enabling stringent constraints on beyond-GR couplings \cite{Crescimbeni:2024sam}. This strategy is particularly promising in scenarios where standard QNM deviations are suppressed but extra modes are present (e.g., in dynamical Chern-Simons gravity), and can naturally be extended to higher harmonics or to specific modified-gravity theories. 

\newpage

\section*{Acknowledgements}

I would like to thank my collaborators Xisco Jimenez-Forteza, Swetha Bhagwat, Julian Westerweck, and Paolo Pani. I acknowledge the financial support provided under the ``Progetti per Avvio alla Ricerca Tipo 1,'' protocol number AR12419073C0A82B. Numerical computations were performed at the Vera cluster, supported by MUR and Sapienza University of Rome.

\bibliography{main}

\end{document}